\documentclass[12pt]{iopart}

\usepackage{iopams}  

\usepackage{amsfonts}
\usepackage{graphicx}
\usepackage{booktabs}
\usepackage{url}
\usepackage[justification=centering]{caption}
\usepackage{multirow}

\begin{document}

\title{PowerSINDy: Identifying Nonlinear Time-Dependent Dynamics in Power Grid Frequency 
}

\author{Xinyi Wen$^1$, Xiao Li$^2$, Leonardo Rydin Gorjão$^{3,4}$, Veit Hagenmeyer$^1$ and Benjamin Schäfer$^1$\thanks{Corresponding author}}

\address{$^1$ Karlsruhe Institute of Technology, Hermann-von-Helmholtz-Platz 1, 76344 Eggenstein-Leopoldshafen, Germany}
\address{$^2$ Electric Power Research Institute Guizhou Power Grid Co., Ltd, China}
\address{$^3$ Department of Environmental Sciences, Faculty of Science, Open University, 6419 AT Heerlen, Netherlands}
\address{$^4$ Faculty of Science and Technology, Norwegian University of Life Sciences, 1430 Ås, Norway}

\ead{xinyi.wen@kit.edu, benjamin.schaefer@kit.edu}
\vspace{10pt}

\begin{abstract}
System identification plays a crucial role in physics and machine learning for discovering governing equations directly from data. A powerful approach is the Sparse Identification of Nonlinear Dynamics (SINDy) method, which assumes that only a few dominant terms drive the essential behavior of a nonlinear dynamical system.
While SINDy methods have shown excellent results, they are most often illustrated on synthetic or simulated systems, leaving open the question of how well they perform on complex, noisy, real-world data. 
Power grid frequency dynamics provide a highly relevant and challenging environment for advancing system identification methods. 
In this work, we propose PowerSINDy as a framework for empirical power system data. 
We apply this framework to empirical frequency data from the Continental Europe (CE) and South Korea (SK) synchronous grids, two major power systems with distinct dynamical characteristics. 
PowerSINDy, which also includes time-dependent terms, can identify the dynamics of these complex real-world systems.
Furthermore, we benchmark three sparsity-promoting regression strategies: Sequentially Thresholded Least Squares (STLSQ), Least Absolute Shrinkage and Selection Operator (LASSO), and Sparse Relaxed Regularized Regression (SR3) to evaluate trade-offs between accuracy, sparsity, and robustness. Results show that LASSO consistently achieves the lowest stable RMSEs, reaching 0.0101 for the CE, while STLSQ provides the best balance between accuracy and stability. SR3 exhibits higher variability and sensitivity to regularization, with $L_0$ and $L_1$ producing nearly indistinguishable outcomes.
\end{abstract}

%
%
%
%
%

\section{Introduction}
The global energy system is undergoing a rapid transformation toward a more sustainable future. Amid this transition, the functioning of modern societies still hinges on the stable operation of electrical power systems. One of the most important metrics for assessing this stability is the power grid frequency, which directly reflects the balance between generation and demand~\cite{machowski2020power}. 
In normal operation, frequency remains close to its nominal reference value (50\,Hz or 60\,Hz, depending on the region), with deviations signaling imbalances in the system~\cite{rydin_gorjao_open_2020,eu-reg-2016/631}. Such deviations induce costly control actions and may affect the reliability of industrial processes and electronic devices~\cite{kundur2022power}.

The efficient operation of power grid systems depends on the accurate characterization of frequency dynamics, which are essential for maintaining system stability and preventing large-scale disturbances. Traditional modeling approaches, most prominently the aggregated swing equation \cite{kundur2022power}, provide a physically intuitive representation of the balance between generation and demand. While such models capture fundamental aspects of system inertia, damping, and stochastic fluctuations, their reliance on linearity and simplifications renders them increasingly insufficient for describing the dynamics of modern grids. The growing penetration of renewable energy sources introduces intermittency, reduced inertia, and nonlinear interactions, all of which demand more flexible and data-driven methods of system identification. Furthermore, power systems are not fully autonomous but driven by load curves and generation schedules, often leading to deterministic signals~\cite{9632335,4840180}. 

This is where system identification plays a crucial role. Bottom-up modeling approaches require detailed knowledge of system parameters, which is often not available in practice. Inference methods, by contrast, offer a complementary top-down approach that allows us to uncover governing dynamics directly from data~\cite{ljung1998system}. For power systems, recent work has already shown the potential of system identification techniques in extracting dynamic models from wide-area frequency measurements~\cite{stiasny2021physics}.

Especially, advances in machine learning and system identification have stimulated the development of methods that infer governing equations directly from time series data. Among these, the Sparse Identification of Nonlinear Dynamics (SINDy) \cite{brunton_discovering_2016, fasel2021sindy, fasel_ensemble-sindy_2022, zolman2024sindy} has attracted significant attention due to its ability to produce parsimonious and interpretable models. By constructing a library of candidate functions and applying sparse regression to select only the most relevant terms, SINDy provides insight into the dominant mechanisms shaping the behavior of complex dynamical systems. While SINDy has been successfully applied across fields such as fluid mechanics~\cite{fukami2021sparse}, biology~\cite{mangan_inferring_2016}, and physics~\cite{corbetta_application_2020}, its application to power system frequency dynamics remains limited~\cite{callaham_nonlinear_2021}. A recent study~\cite{wen2025identifying} presented a proof-of-concept application of SINDy to complex empirical frequency data from multiple synchronous grids, demonstrating that the method could uncover nonlinear and time-dependent dynamics in noisy, real-world datasets. However, this initial study did not systematically explore how optimizer choice, or feature complexity, affects model accuracy and stability.

%
Applying SINDy to grid frequency data presents challenges. Oscillatory dynamics in frequency recordings, such as damped modes or periodic fluctuations, cannot be well represented by polynomial bases alone. Although SINDy can incorporate trigonometric functions, power grid frequency oscillations are not strictly periodic and are driven by stochastic and deterministic disturbances. This leads to sparse regression being unstable~\cite{kundur2022power, callaham2022role, fasel2021sindy}.
As a result, choosing an appropriate optimizer and regularization strategy becomes critical for obtaining a reliable and interpretable sparse model.
In this study, we propose the PowerSINDy framework tailored to the characteristics of power system frequency dynamics and other time-dependent systems. The PowerSINDy framework extends the candidate library to include quadratic and cubic polynomial terms as well as Fourier functions, thereby enhancing the representation of nonlinear couplings and oscillatory behaviors. Furthermore, we benchmark three different sparsity-promoting regression strategies—Sequentially Thresholded Least Squares (STLSQ), SINDy with LASSO~\cite{ranstam2018LASSO}, and Sparse Relaxed Regularized Regression (SR3)~\cite{champion_unified_2020}-to systematically evaluate the trade-offs between model accuracy, stability, and interpretability.

We apply the PowerSINDy framework to high-resolution frequency data collected from two synchronous grids, Continental Europe (CE) and South Korea (SK). After careful preprocessing and Gaussian smoothing to isolate deterministic dynamics from stochastic fluctuations, we identify governing equations under different model configurations and optimization strategies. Our analysis compares results across datasets, investigates the influence of polynomial order and Fourier augmentation, and quantifies the structural complexity of the discovered models. In doing so, the study provides new insights into the capabilities and limitations of data-driven identification in capturing the nonlinear and oscillatory nature of power grid frequency dynamics.

The remainder of the paper is organized as follows. Section~\ref{sec:frequency data preprocessing} introduces the preprocessing pipeline and smoothing optimization for frequency data. Section~\ref{sec:methodology} presents the PowerSINDy framework and sparsity-promoting regression strategies. Section~\ref{sec:results} reports empirical results across datasets, configurations, and optimizers. Finally, Section~\ref{sec:conclusion} summarizes the findings and outlines directions for future work.

\section{Frequency Data Preprocessing and Smoothing Optimization}\label{sec:frequency data preprocessing}

We collected frequency measurements from two complementary synchronous power grids as test cases for PowerSINDy: the Continental Europe (CE) grid and the South Korea (SK) grid. The CE grid operates at a nominal frequency of 50\,Hz, while the SK grid operates at a nominal frequency of 60\,Hz. Both datasets were recorded at 1-second resolution. The CE grid is one of the largest interconnected power systems in the world, supplying electricity to hundreds of millions of consumers across 32 countries with a generation capacity of approximately 800 GW~\cite{entsoe_ucte_annual_report_2008}. This massive scale and interconnectedness create a system with enormous rotational inertia, primarily due to the thousands of large synchronous generators connected in the network.
In contrast, the SK grid is a geographically compact, nationally operated system with an installed generation capacity of roughly 144\,GW~\cite{reglobal_sk_2024}.
This compactness results in a system with lower total rotational inertia compared to CE. Unlike the multi-country governance of the CE grid, SK's system is centrally managed, allowing for highly coordinated control actions.
We analyzed 15-minute trajectories at 1-second resolution to capture both the market scale (minutes to hours) and the dynamic stability of these systems (seconds). The CE data span the period from January 1, 2024 to April 24, 2024, while the SK data cover the period from August 15, 2024, to December 10, 2024~\cite{power_grid_frequency}. 
SK dataset is collected by an electrical data recorder (EDR)~\cite{jumar2020database, powertech}, which is a monitoring and analysis tool for electrical power grids that continuously collects and stores comprehensive measurement data in the Karlsruhe Institute of Technology (KIT) large-scale database for subsequent processing.
To ensure data integrity, we excluded rows containing missing values and retained only complete 15-minute segments, each comprising 900 samples. This filtering process resulted in 10,939 valid 15-minute records for CE and 10,405 valid 15-minute records for SK, which were used in subsequent analysis.

Before applying the SINDy algorithm~\cite{brunton_discovering_2016}, we employed a Gaussian filter to reduce noise while preserving the intrinsic oscillatory characteristics of the angular frequency data. We randomly selected 10 days from the available valid dates, focusing on the 9 AM time point. For each selected day, we evaluated a range of $\sigma$ values, where $\sigma$ denotes the bandwidth of the Gaussian filter, controlling the degree of smoothing. We applied the filter to the corresponding frequency values, generating a smoothed angular frequency for each $\sigma$ value.

\begin{figure}
\centering
\includegraphics[width=1\columnwidth]{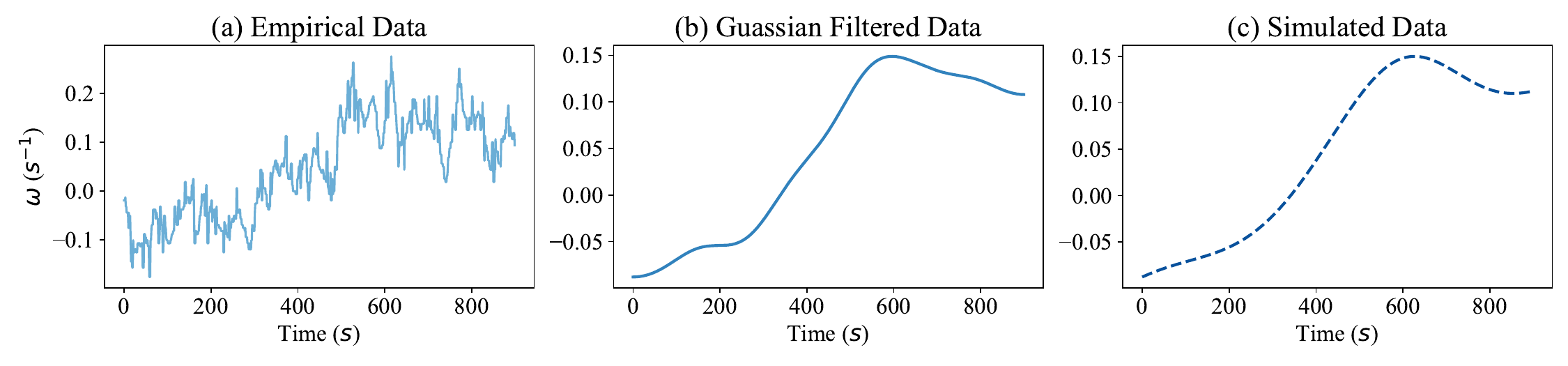}
\caption{\label{fig:combined}Schematic of the dynamics discovery pipeline. (a) Raw empirical frequency data. (b) The data are preprocessed using a Gaussian filter, where the optimal parameter $\sigma$ is determined by minimizing the RMSE. (c) The smoothed data is then analyzed using the SINDy algorithm to discover the governing equations.
}
\end{figure}

To determine the optimal smoothing parameter, we evaluated the performance of Gaussian filtering across different $\sigma$ values by calculating the root mean square error (RMSE) between SINDy simulated trajectories and the original empirical data. For each $\sigma$ value, we applied the Gaussian filter to the angular velocity ($\omega$) measurements, identified the underlying dynamics using SINDy, simulated the system, and computed the RMSE between the simulated and empirical time series. The $\sigma$ value yielding the lowest RMSE was identified as the optimal smoothing parameter, balancing noise suppression with preservation of essential dynamical features. 
The process is demonstrated in Figure~\ref{fig:combined}.

\begin{figure}
\centering
\includegraphics[width=1\columnwidth]{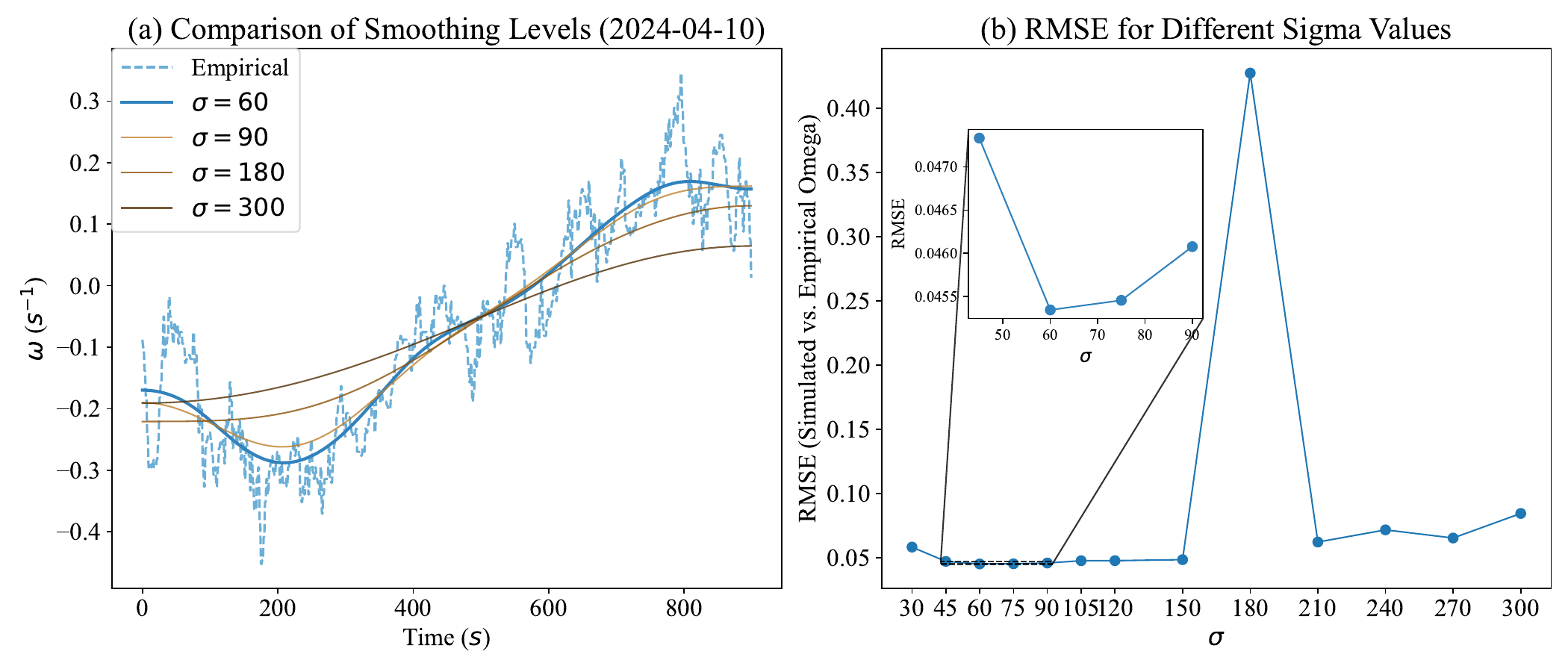}
\caption{\label{fig:different_sigma} Comparison of smoothing levels and corresponding model performance. (a) Time series of angular velocity ($\omega$) showing empirical data (blue dashed line) and Gaussian-filtered trajectories at different smoothing parameters ($\sigma$ = 60, 90, 180, 300). The $\sigma$ = 60 filter (blue solid line) provides optimal balance between noise reduction and feature preservation. (b) RMSE between SINDy simulated and empirical $\omega$ as a function of smoothing parameter $\sigma$. The inset highlights the optimal range ($\sigma$ = 45--90) where RMSE reaches its minimum value of 0.0453 at $\sigma$ = 60, indicating this smoothing level yields the most accurate dynamical model.
}
\end{figure}

Our analysis revealed that $\sigma = 60$ (corresponding to a 60$\mathrm{s}$ window at 1 Hz sampling) achieved the minimum RMSE of 0.0453, indicating superior model performance at this smoothing level (Figure~\ref{fig:different_sigma}). The RMSE remained relatively stable $\sigma$ values between 45 and 90, while substantially increasing beyond 180 due to oversmoothing that eliminated critical dynamical information. Based on these findings, we adopted $\sigma = 60$ as the optimal smoothing parameter and applied it consistently to both CE frequency measurements and SE grid data. This standardized preprocessing approach ensures comparable conditions for subsequent nonlinear dynamical analysis across both synchronous grid systems.



\section{Methodology}\label{sec:methodology}
\subsection{Data-Driven Model}

The basis for modeling the aggregate frequency dynamics of a synchronous power grid is the swing equation \cite{kundur2022power}, which describes the motion of a synchronous machine subject to damping, restoring forces, and stochastic disturbances. At the aggregated system level, the dynamics of the grid frequency can be approximated by a linear stochastic differential equation of the form~\cite{Schäfer2018non,wen2025identifying, oberhofer}
\begin{eqnarray}
\frac{\mathrm{d}\theta}{\mathrm{d}t} &= \omega,  \\
\frac{\mathrm{d}\omega}{\mathrm{d}t} &= -c_{\omega}\omega - c_{\theta}\theta + \Delta P(t) + \epsilon\xi(t),
\end{eqnarray}
where $\theta$ denotes the bulk angle deviation, $\omega = 2\pi(f-f_{\mathrm{ref}})$ the frequency deviation from the nominal reference $f_{\mathrm{ref}}=50$\,Hz for CE data and $f_{\mathrm{ref}}=60$\,Hz for SK data, $c_\omega$ the damping coefficient, and $c_\theta$ the stiffness coefficient associated with secondary control. The input $\Delta P(t)$ represents the power imbalance between generation and demand, while $\epsilon \xi(t)$ models stochastic fluctuations via a Gaussian white noise process $\xi(t)$ with amplitude $\epsilon$.

This aggregated swing equation captures the essential physical intuition of power grid dynamics: frequency deviations are driven by imbalances between generation and demand, damped by primary and secondary control mechanisms, and perturbed by stochastic fluctuations. Comprehensive discussions of the swing equation, its parameterizations, and common modeling approaches can be found in~\cite{gorjao_data-driven_2020,oberhofer}. However, this model remains limited in scope, as it assumes linearity and does not account for nonlinear interactions, higher-order couplings, or oscillatory modes that arise in large-scale synchronous grids. To overcome these limitations, we employ a novel data-driven approach called the PowerSINDy framework. The implementation of the PowerSINDy framework is publicly available on GitHub \cite{wen_github}.

Prior to model identification, the frequency data were smoothed using a Gaussian filter to suppress measurement noise while preserving intrinsic oscillatory behavior. As a result, we assume that residual stochastic fluctuations are negligible, allowing the system to be treated as effectively deterministic for regression purposes.

Under this assumption, SINDy identifies the deterministic dynamics of the filtered angular frequency and phase as

\begin{eqnarray}
\frac{\mathrm{d}\theta}{\mathrm{d}t} &=& \omega,  \\
\frac{\mathrm{d}\omega}{\mathrm{d}t} &=& f(\theta, \omega,T),
\end{eqnarray}
where the function $f(\theta, \omega,T)$ represents a sparse combination of candidate terms drawn from the library $\Theta(\mathbf{X})$ (e.g., polynomial, cross, or higher-order interaction terms with time-dependent), see Eq.\ref{eq3}. This formulation generalizes the classical linear swing equation by allowing SINDy to uncover possible nonlinear higher-order terms.

\subsection{Sparse Regression Formulations}

The identification of governing equations in the SINDy framework reduces to a sparse regression problem of the form
\begin{equation}
\dot{\mathbf{X}} = \Theta(\mathbf{X}) \Xi,
\label{eq3}
\end{equation}
where $\mathbf{X}$ is the state matrix, $\dot{\mathbf{X}}$ is its time derivative, $\Theta(\mathbf{X})$ is the candidate function library, and $\Xi$ is the coefficient matrix to be estimated. To promote sparsity in $\Xi$, we employ three complementary optimization strategies~\cite{brunton_discovering_2016}.

\subsubsection{Sequentially Thresholded Least Squares (STLSQ):}

STLSQ iteratively solves a least-squares problem and eliminates small coefficients below a user-defined threshold $\lambda$. The optimization problem for STLSQ is formulated as:


\begin{equation}
\hat{\Xi} = \mathop{\mathrm{argmin}}_{\Xi} \; 
\left\| \Theta(X)\Xi - \dot{X} \right\|_2^2 
+ \alpha \left\| \Xi \right\|_2.
\end{equation}
Here, $\alpha$ is a ridge-type regularization parameter penalizing large coefficients.
After each iteration, coefficients with magnitudes smaller than the threshold $\lambda$ are set to zero:

\begin{equation}
\Xi_{ij} \leftarrow 0 \quad \mbox{if} \quad |\Xi_{ij}| < \lambda.
\end{equation}
while $\lambda$ denotes the threshold applied to enforce sparsity. After solving the regression, coefficients with magnitude smaller than $\lambda$ are set to zero in an iterative refinement procedure. This promotes sparsity while maintaining computational efficiency.

The hyperparameters $\lambda$ and $\alpha$ are selected via a systematic grid search (See ~\ref{app:hyperparam}), exploring $\lambda$ over multiple orders of magnitude and $\alpha$ over a wide regularization range. The values reported in the Results section correspond to configurations that achieve a robust balance between sparsity and accuracy.

\subsubsection{LASSO-SINDy:}

LASSO regression introduces an $L_{1}$ penalty into the optimization, directly promoting sparsity through coefficient shrinkage~\cite{ranstam2018LASSO}:

\begin{equation}
\hat{\Xi} = \mathop{\mathrm{argmin}}_{\Xi} \;  
\left\| \Theta(X)\Xi - \dot{X} \right\|_2^2 
+ \alpha \left\| \Xi \right\|_1.
\end{equation}

In this formulation, $\alpha$ is the $L_{1}$ regularization weight that shrinks many coefficients exactly to zero, yielding sparse and interpretable governing equations. 
The parameter $\alpha$ is determined via grid search~(\ref{app:hyperparam}).

\subsubsection{Sparse Relaxed Regularized Regression (SR3):}

SR3 reformulates the sparse regression problem by introducing an auxiliary variable $W$ and a relaxation parameter $\nu$, reformulating the regression problem as


\begin{equation}
\hat{\Xi}, \hat{W} = \mathop{\mathrm{argmin}}_{\Xi,\,W} \; 
\frac{1}{2}\left\| \Theta(X)\Xi - \dot{X} \right\|_2^2 
+ \kappa R(W) 
+ \frac{\nu}{2} \left\| \Xi - W \right\|_2^2.
\end{equation}

where $R(W)$ is a sparsity-promoting penalty, we employed three different norms, $L_0$, $L_1$ and $L_2$. A key idea in SR3 is the relaxation term $\frac{\nu}{2}\|\Xi - W\|_2^2$, which allows $\Xi$ to focus on fitting the data, while $W$ handles the sparsity.  This separation often enhances robustness to noise and convergence properties. The parameter $\kappa$ controls the strength of the spasity penalty, while $\nu$ determines the strength of relaxation between $\Xi$ (fitting the data) and $W$ (handling sparsity)~\cite{champion_unified_2020}.  

 A key feature of SR3 is the relaxation mechanism, which balances between solution sparsity and regression accuracy, improving robustness in the presence of measurement noise and convergence behavior.

Hyperparameters $\kappa$ and $\nu$ are selected by grid search~(\ref{app:hyperparam}).
The values used in the Results section differ between datasets. A larger value ($\nu=10$) is required for the SK dataset to achieve stable convergence, as lower values (e.g., $\nu = 1$) failed to converge within the iteration limits and degraded model performance.

\subsection{Introducing PowerSINDy for time-dependent and non-polynomial dynamics}

To extract interpretable models of the deterministic frequency dynamics, we adopted a modified version of the SINDy framework. The central idea of SINDy is to construct a library of candidate functions from the observed time series and then identify a parsimonious subset of terms that best represents the governing equations. In this work, we extend the standard polynomial library in two directions~\cite{fung2025rapid, lee2024energy,hosseinipour2023data}. First, we incorporate quadratic and cubic terms ($p=2,3$) to capture nonlinear interactions between state variables. Second, we augment the quadratic library with Fourier terms, enabling the framework to represent oscillatory components that naturally arise in synchronous power grid dynamics due to generation-load imbalances.
Throughout the following Results Section, we use the shorthand notation \(p2\), \(p3\), and \(p2f1\) to denote different choices of the candidate function library. Specifically, \(p2\) corresponds to a quadratic polynomial library, \(p3\) extends this to cubic polynomials, and \(p2f1\) denotes a quadratic polynomial library augmented with first-order Fourier terms. For the \(p2f1\) library used in this study, the features included are:
$\left\{1, \theta, \omega, T, \theta^2, \theta\omega, \theta T, \omega^2, \omega T, T^2, \sin(\theta), \cos(\theta), \sin(\omega), \cos(\omega), \sin(T), \cos(T)\right\}.$

We do not consider higher-order combinations such as \(p3f1\) or \(p2f2\), since preliminary tests already indicated a tendency toward overfitting when increasing polynomial order, and further enlarging the library would substantially increase computational cost without improving predictive performance. This restriction allows us to keep the function space expressive but still parsimonious.

Polynomial terms capture nonlinear couplings between dynamical states, which are particularly relevant in large-scale interconnected grids where disturbances can propagate nonlinearly across subsystems. Fourier terms, on the other hand, provide a representation of oscillatory behaviors, which are poorly approximated by polynomials alone. The inclusion of Fourier functions thus enhances the ability of the novel PowerSINDy framework to capture both transient and sustained oscillations in frequency dynamics.

To robustly determine the governing equations, we apply the three sparse regression strategies described above: STLSQ, LASSO, and SR3. Each optimizer imposes sparsity through a different mechanism, thereby providing complementary perspectives on model selection. By systematically comparing the identified models across these optimizers and function libraries, we aim to assess both the robustness and interpretability of the governing equations derived from real-world power grid frequency data.



\section{Results}\label{sec:results}

\subsection{Dataset-level comparison: CE vs.\ SK}
Across all configurations and optimizers, the CE dataset tends to yield slightly lower RMSE values than the SK dataset, while stability fractions (the percentage of trajectory chunks for which the identified model produced non-divergent dynamics) remain broadly comparable. For example, in Table~\ref{tab:ce-sk-comparison}, under LASSO with the \(p2\) library, CE achieves a mean stable RMSE of \(0.0163\) (the root mean squared error averaged only over those stable chunks), closely matching SK at \(0.0161\). For higher-order models, the gap becomes more evident: in the \(p3\) case, CE significantly outperforms SK (\(0.0101\) vs.\ \(0.0120\)), indicating that CE data may facilitate better recovery of complex dynamics. Stability fractions, however, are generally similar across datasets, with both regions achieving values around \(0.68\)–\(0.84\) depending on optimizer and configuration. This suggests that differences in RMSE likely reflect intrinsic data characteristics (e.g. easier, more predictable dynamics) rather than algorithmic instability.


\subsection{Identifying the best candidate functions}
Model complexity exerts a notable influence on performance. The quadratic polynomial (\(p2\)) models generally produce stable solutions with low RMSE, particularly under LASSO, where both datasets achieve errors near \(0.0160\). When the polynomial degree is increased to three (\(p3\)), CE benefits more strongly than SK, yielding its best overall result with LASSO (\(0.0101\)) while SK achieves only \(0.0120\). However, this gain in accuracy comes at the cost of reduced stability, especially for SR3 on SK (\(\approx 0.45\)–\(0.51\)). 
This behavior arises because higher-order polynomial terms lead to faster divergence of the reconstructed trajectories toward the end of the prediction interval. As a result, cubic models are harder to fit robustly and exhibit a larger proportion of diverging, i.e. unstable, trajectories.
The Fourier-augmented models (\(p2f1\)) provide an middle ground: they increase representational capacity without destabilizing the optimization. Both CE and SK achieve very similar performance in this setting, with RMSE around \(0.016\) and stability fractions of \(0.67\)–\(0.81\). These results suggest that including Fourier modes may help mitigate regional differences, producing robust and balanced models across datasets.

\subsection{Choosing the best optimizer}
The three optimizers exhibit markedly different behaviors in terms of accuracy and stability, see Table~\ref {tab:ce-sk-comparison}. LASSO achieves the lowest stable RMSE across datasets and model configurations, with values such as \(0.0161\) for SK–\(p2f1\) and \(0.0160\) for CE–\(p2f1\). Particularly in the \(p3\) case, CE under LASSO attains an RMSE of \(0.0101\), representing the best performance observed across all experiments. Stability fractions under LASSO are moderately high, typically ranging between \(0.52\) and \(0.84\), but they do not consistently reach the highest levels. By contrast, STLSQ emerges as the most robust method with respect to stability. In both SK and CE datasets, stability fractions under STLSQ regularly exceed \(0.79\) in the \(p2\) setting, and remain close to \(0.75\) in the \(p2f1\) case. This robustness is achieved with only a modest increase in RMSE compared to LASSO, suggesting that STLSQ offers the most favorable trade-off between accuracy and reliability. SR3 presents the most heterogeneous performance profile. In the \(p2\) setting, CE under SR3 reaches low RMSE values (\(0.0156\)) with stability fractions around \(0.68\), while SK produces slightly higher RMSE (\(0.0170\)–\(0.0220\)) with comparable stability (\(0.66\)–\(0.70\)). However, the situation changes drastically for the \(p3\) configuration, where SK stability fractions fall below \(0.52\), reflecting a considerable deterioration. These findings indicate that SR3 is highly sensitive to the choice of norm, hyperparameters, and dataset characteristics, making it less reliable than LASSO or STLSQ in scenarios demanding consistent performance

It is worth noting that, for SR3, both the $L_0$ and $L_1$ norms have indistinguishable performance in terms of both RMSE and stability fraction. This suggests that, under the present setting, the $L_0$ and $L_1$ penalties become effectively equivalent.

\begin{figure}
\centering
\includegraphics[width=1\columnwidth]{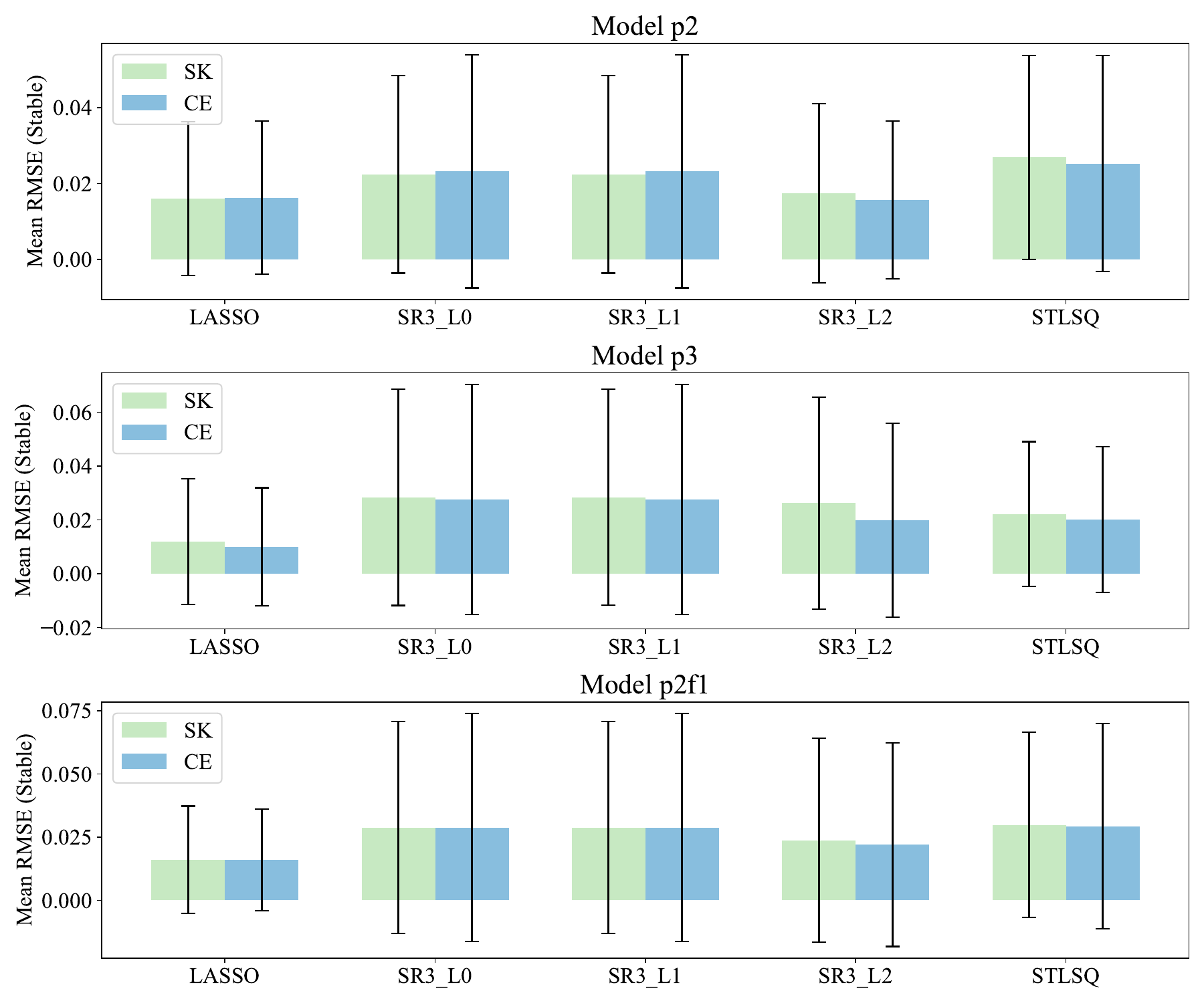}
\caption{\label{fig:optrmseerrorbars} 
Comparison of mean stable RMSE with standard deviation error bars across optimizers and regularization types. Error bars indicate the variability of RMSE across stable chunks. LASSO consistently achieves the lowest mean RMSE with minimal variance, reflecting its strong numerical stability. STLSQ maintains a favorable balance between accuracy and consistency, while SR3 exhibits greater variability, highlighting its sensitivity to regularization choice and dataset characteristics.
}
\end{figure}

\begin{figure}
\centering
\includegraphics[width=1\columnwidth]{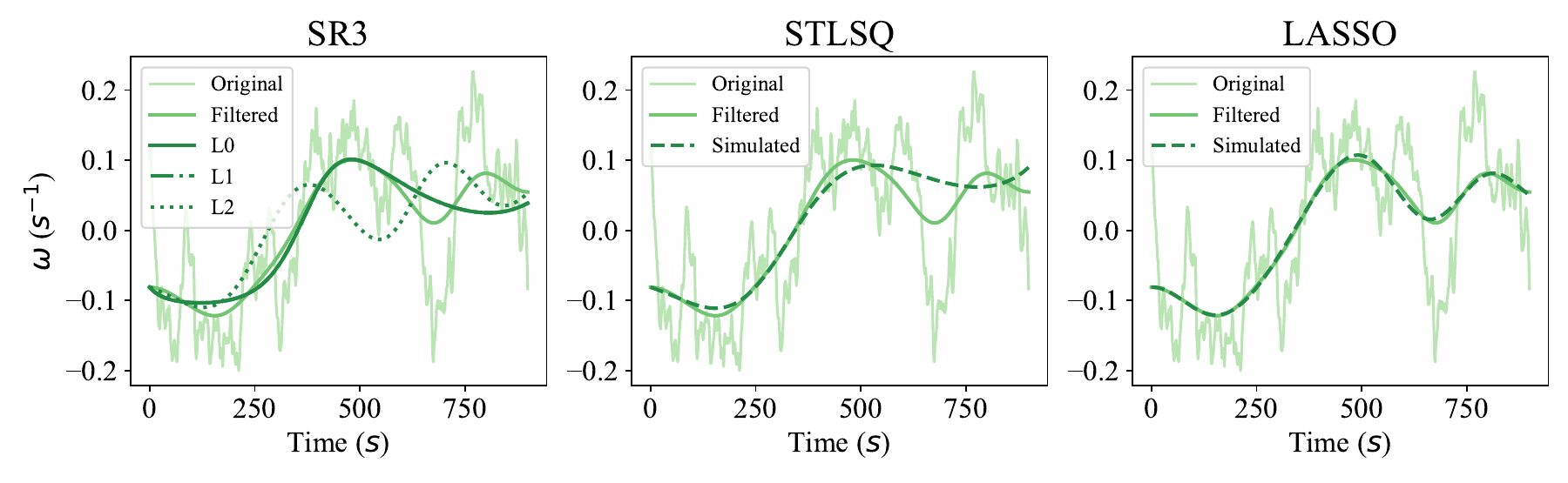}
\caption{\label{fig:opt} 
Comparison of simulated $\omega(t)$ trajectories using SR3, STLSQ, and LASSO. SR3 results are shown for different regularization norms, illustrating the effect of regularization choice. STLSQ and LASSO reconstructions demonstrate higher agreement with the filtered signal, particularly LASSO, which achieves the closest alignment to the original dynamics.
}
\end{figure}

The error bar analysis in Figure~\ref{fig:optrmseerrorbars} further substantiates these performance patterns, revealing distinct variability characteristics across optimization methods. LASSO demonstrates not only superior mean accuracy but also exhibits the most consistent performance with the smallest error variability across different operational conditions. STLSQ maintains comparably narrow error bars, reflecting its strong stability, albeit with slightly higher mean RMSE. In contrast, Conversely, SR3 methods, particularly in complex model configurations, show substantially larger error bars, indicating higher performance instability and sensitivity to varying system conditions. An example of the trajectory comparison in Figure~\ref{fig:opt} highlights the varying simulation fidelity among optimization methods. Notably, LASSO delivers the most accurate performance, with its simulated trajectories nearly indistinguishable from the Gaussian filtered signal, thereby confirming its superior balance of numerical stability and predictive accuracy under sparse regularization constraints. 

To further contextualize the performance of the sparse identification models, we compare them against a baseline reference generated by the Euler–Maruyama integration of the underlying stochastic differential equations. The empirical RMSE between the real and baseline-simulated frequency trajectories amounts to \(0.0282\)  for the CE dataset and 
\(0.0342\) for the SK dataset. While these values are relatively low, they remain consistently higher than those achieved by the best-performing SINDy configurations (e.g., LASSO and STLSQ). This comparison highlights that sparse model discovery approaches reconstruct the system dynamics with greater accuracy than the stochastic integration baseline. Consequently, the SINDy frameworks demonstrate superior data efficiency and predictive precision, confirming their advantage in reconstructing governing dynamics from noisy empirical signals.

\begin{table}[ht]
\centering
\caption{Comparison of mean stable RMSE and stability fraction for CE and SK datasets across models, optimizers, and configurations. For reference, the baseline simulation yields RMSE values of \(0.0282\) (CE) and \(0.0342\) (SK). All reported configurations correspond to hyperparameters selected from a systematic grid search on the CE dataset (see~\ref{app:hyperparam} for full parameter sensitivity analysis).}
\label{tab:ce-sk-comparison}
\resizebox{0.8\textwidth}{!}{%
\begin{tabular}{lllcccc}
\toprule
 & & & \multicolumn{2}{c}{CE} & \multicolumn{2}{c}{SK} \\

Model & Optimizer & Config & RMSE & Stability & RMSE & Stability \\
\midrule
p2  & LASSO & $\alpha=10^{-6}$ & 0.0163 & \textbf{0.844} & \textbf{0.0161} & \textbf{0.829} \\
    & SR3 ($L_0$)   &  $\kappa=10^{-6}, \nu=1.0$ & 0.0232 & 0.677 & 0.0224 & 0.703 \\
    & SR3 ($L_1$)   &  $\kappa=10^{-6}, \nu=1.0$ & 0.0232 & 0.677 & 0.0224 & 0.703 \\
    & SR3 ($L_2$)   &  $\kappa=10^{-6}, \nu=1.0$ & \textbf{0.0156} & 0.681 & 0.0175 & 0.658 \\
    & STLSQ & $\lambda=10^{-6}, \alpha=10$   & 0.0253 & 0.797 & 0.0269 & 0.793 \\
\midrule    
p2f1& LASSO & $\alpha=10^{-6}$ & \textbf{0.0160} & \textbf{0.814} & \textbf{0.0161} & \textbf{0.795} \\
    & SR3 ($L_0$)   & $\kappa=10^{-6}, \nu=1.0$ & 0.0288 & 0.669 & 0.0287 & 0.677 \\
    & SR3 ($L_1$)   & $\kappa=10^{-6}, \nu=1.0$ & 0.0288 & 0.669 & 0.0287 & 0.677 \\
    & SR3($L_2$)  & $\kappa=10^{-6}, \nu=1.0$ & 0.0220 & 0.667 & 0.0238 & 0.634 \\
    & STLSQ & $\lambda=10^{-6}, \alpha=10$   & 0.0293 & 0.747 & 0.0299 & 0.746 \\
\midrule    
p3  & LASSO & $\alpha=10^{-6}$ & \textbf{0.0101} & 0.586 & \textbf{0.0120} & 0.525 \\
    & SR3 ($L_0$)   &  $\kappa=10^{-6}, \nu=1.0$ & 0.0276 & 0.538 & --     & --    \\
    & SR3 ($L_0$)  &  $\kappa=10^{-6}, \nu=10$  & --     & --    & 0.0284 & 0.515 \\
    & SR3 ($L_1$)   & $\kappa=10^{-6}, \nu=1.0$ & 0.0276 & 0.538 & --     & --    \\
    & SR3 ($L_1$)  &  $\kappa=10^{-6}, \nu=10$  & --     & --    & 0.0284 & 0.515 \\
    & SR3 ($L_2$)  & $\kappa=10^{-6}, \nu=1.0$ & 0.0199 & 0.518 & --     & --    \\
    & SR3 ($L_2$)  &  $\kappa=10^{-6}, \nu=10$  & --     & --    & 0.0262 & 0.454 \\
    & STLSQ & $\lambda=10^{-6}, \alpha=10$   & 0.0202 & \textbf{0.639} & 0.0222 & \textbf{0.658} \\
\bottomrule
\end{tabular}
}
\end{table}

\subsection{Obtaining sparse solutions}
\begin{figure}
\centering
\includegraphics[width=1\columnwidth]{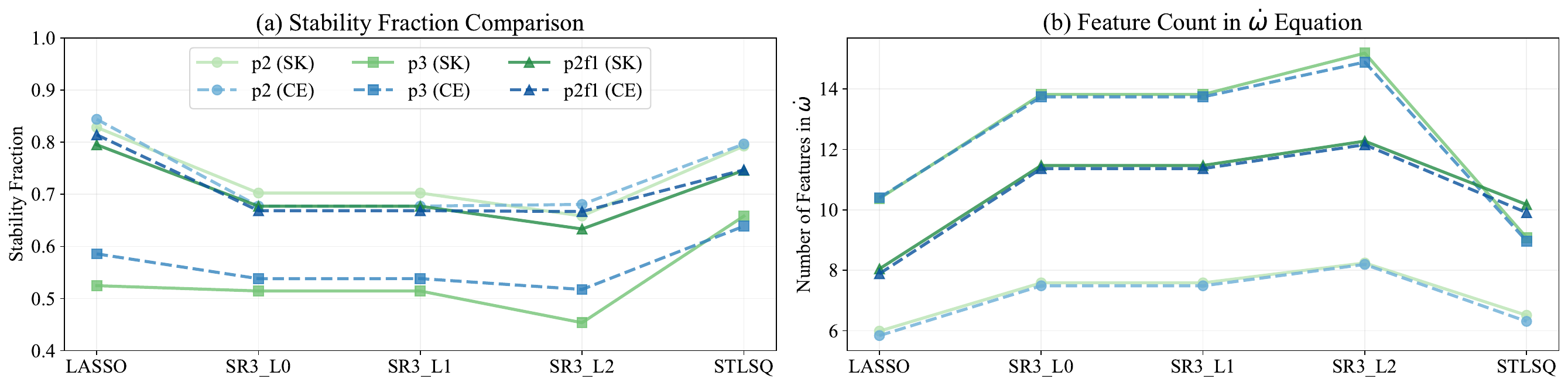}
\caption{\label{fig:stability_and_features} 
Comparative analysis of model stability and feature selection: (a) Stability fraction across different models and optimizers for both SK and CE grid systems. Solid lines with filled markers represent SK grid data, while dashed lines with open markers denote CE grid data. (b) Number of identified features in the $\dot{\omega}$ equation under threshold $1e-6$, illustrating the sparsity characteristics of each optimization approach. 
}
\end{figure}

Beyond accuracy and stability, we investigated the structural complexity of the discovered models by quantifying the number of active coefficients in the governing equations. Figure~\ref{fig:stability_and_features} reports the average number of features identified under a threshold of \(10^{-6}\). 

For the \(p2\) models, all optimizers converged to relatively parsimonious representations, with SK and CE consistently requiring between 6 and 8 active features. LASSO produced the sparsest models, with mean counts of approximately \(5.9\) for both datasets, while  SR3 and STLSQ retained slightly more terms ($\approx 7.7$ and $\approx 6.4$, respectively). Importantly, the differences between SK and CE in this setting were minimal, suggesting that the learned feature complexity is largely invariant to dataset choice when the polynomial order is limited to two.

The $n_2f_1$ case exhibited moderately increased complexity due to the inclusion of additional nonlinear interactions. SR3 and STLSQ identified 10--12 active features on average, whereas LASSO remained the most parsimonious, converging to $\approx 8$ features. These results indicate that LASSO continues to enforce effective sparsity even as model richness increases.

The situation changes markedly in the \(p3\) setting, where model dimensionality naturally increases. SR3 consistently produced the densest structures with $\approx 14$ active terms, STLSQ identified around 9, and LASSO maintained a moderate level of sparsity ($\approx 10$). This demonstrates that SR3 tends to favor richer polynomial representations, while LASSO maintains controlled sparsity at the cost of slightly higher bias.

Overall, the feature complexity analysis highlights that LASSO systematically enforces sparsity without sacrificing predictive performance, while STLSQ and SR3 tend to favor denser models, particularly as the polynomial order increases. This trade-off between model simplicity and stability is consistent with their respective optimization strategies and provides valuable guidance for selecting regularization approaches in data-driven discovery of dynamics. 
Furthermore, we observed that the nonlinear terms $\theta\omega$ and $\theta\omega^2$ make significant contributions in the LASSO \(p3\) model, as evidenced by their relatively large coefficients, indicating that these interactions play an important role in capturing the system dynamics.

\section{Conclusion and Outlook}\label{sec:conclusion}

This work presented the novel PowerSINDy framework for uncovering governing equations for nonlinear, time-dependent dynamics and validated them on empirical power grid frequency data. We included explicit time-dependency and selected suitable candidate functions, in particular higher-order polynomial and Fourier terms. By systematically comparing sparse regression strategies, we demonstrated the ability of the framework to recover interpretable models from both synthetic and real-world grid data. Experiments on Continental Europe (CE) and South Korea (SK) datasets revealed several key insights. 
Data-level comparisons showed that the CE grid, being geographically extensive and highly interconnected, generally yielded lower RMSE and higher stability fractions, whereas the SK grid, with smaller generation capacity and a more compact topology, presented greater variability.
Optimizer-level comparison showed that LASSO consistently achieved the lowest RMSE values while enforcing compact representations, highlighting its suitability for accurate and parsimonious modeling. STLSQ emerged as the most stable optimizer, particularly for larger or more complex features, providing robust recovery across configurations. In addition, SR3 exhibited highly variable performance, indicating sensitivity to hyperparameters and dataset properties, with $L0$ and $L1$ norms yielding closely comparable outcomes. These findings emphasize the importance of optimizer choice and function library design in data-driven discovery of nonlinear, time-dependent power system dynamics. We also revealed that in the LASSO $p3$ model, the nonlinear interactions $\theta\omega$ and $\theta\omega^2$ contributed most strongly,  indicating their critical role in capturing system dynamics. 

Beyond quantitative accuracy, our feature complexity analysis revealed that higher-order libraries significantly increase the number of active terms, raising questions about interpretability and overfitting. The inclusion of Fourier terms provided a balanced alternative, enabling robust representation of oscillatory modes without excessive model complexity. These observations underscore the trade-off between expressiveness and interpretability that lies at the core of data-driven modeling. 

Looking ahead, several research directions emerge. First, extending the framework to incorporate explicit time-varying parameters could capture the evolving dynamics of renewable-rich grids. Second, online or adaptive SINDy formulations may enable real-time monitoring of stability margins under changing system conditions. Third, integrating prior physical knowledge (e.g., inertia constants or control dynamics) with data-driven discovery could yield hybrid models that combine interpretability with adaptability. Ultimately, such advances hold promise for equipping system operators with interpretable, data-driven models that enhance situational awareness and resilience in modern power grids.

\section{Acknowledgments}
This work was funded by the Helmholtz Association and the Networking Fund through Helmholtz AI and under grant no. VH-NG-1727 as well as by the Deutsche Forschungsgemeinschaft (DFG, German Research Foundation) -- 556503410. 

The authors gratefully acknowledge the computing time provided on the high-performance computer HoreKa by the National High-Performance Computing Center at KIT (NHR@KIT). This center is jointly supported by the Federal Ministry of Education and Research and the Ministry of Science, Research and the Arts of Baden-W\"{u}rttemberg, as part of the National High-Performance Computing (NHR) joint funding program (\url{https://www.nhr-verein.de/en/our-partners}). HoreKa is partly funded by the German Research Foundation (DFG).

\section{References}
\bibliographystyle{iopart-num}
\bibliography{references.bib}

\newpage
\appendix
\section{Hyperparameter Sensitivity Analysis}\label{app:hyperparam}
To ensure a robust and unbiased selection of model hyperparameters, we perform a systematic grid search over all optimization methods on the CE dataset. The explored parameter ranges span several orders of magnitude. 
For Lasso, the sparsity parameter $\alpha$ varies from $10^{-10}$ to $10^{-3}$, together with solver tolerance values $\texttt{tol} \in [10^{-7}, 10^{-6}]$.
The STLSQ optimizer involves a two-dimensional grid over the sparsity threshold $\lambda \in[10^{-10}, 10^{-3}]$ and ridge regularization parameter $\alpha \in[10^{-3}, 10^{1}]$. For SR3, we vary the sparsity regularization parameter $\kappa \in [10^{-10} , 10^{-3}]$, and the relaxation parameter $\nu$ over five values from $10^{-3}$ to $10^{1}$.
The tolerance parameter controls the convergence criterion of the optimization solver. Across all tested configurations where it is applicable, its effect on RMSE and stability is negligible.

\subsection{Sensitivity to Library Complexity}
\begin{figure}
\centering
\includegraphics[width=1.1\columnwidth]{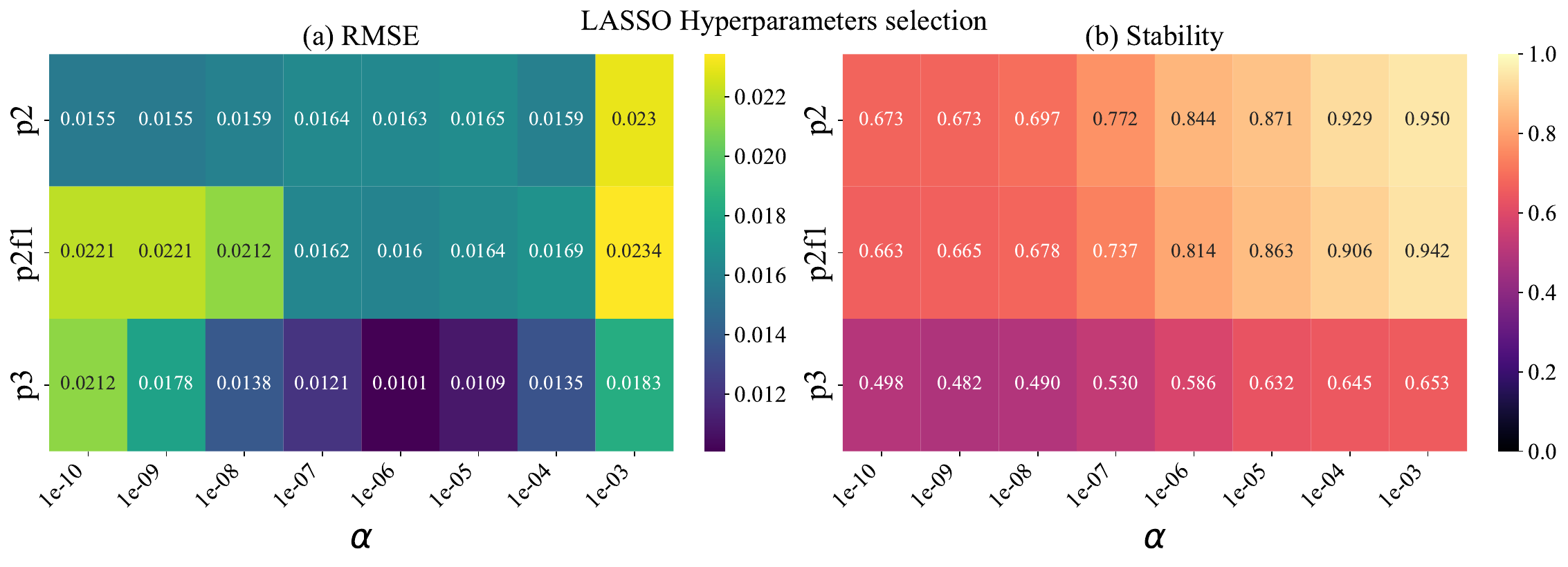}
\caption{\label{fig:heatmap_lasso}Hyperparameter Sensitivity Analysis for LASSO. Heatmaps illustrating the performance of three model configurations across a range of sparsity parameter. (a)RMSE values indicate the accuracy, with the lower error achieved by $p3$ at $\alpha$=$10^{-6}$. (b) Stability fractions indicate the propotion of non-divergent trajectories. Stability increases with $\alpha$ across all configurations.
}
\end{figure}

\begin{figure}
\centering
\includegraphics[width=1.1\columnwidth]{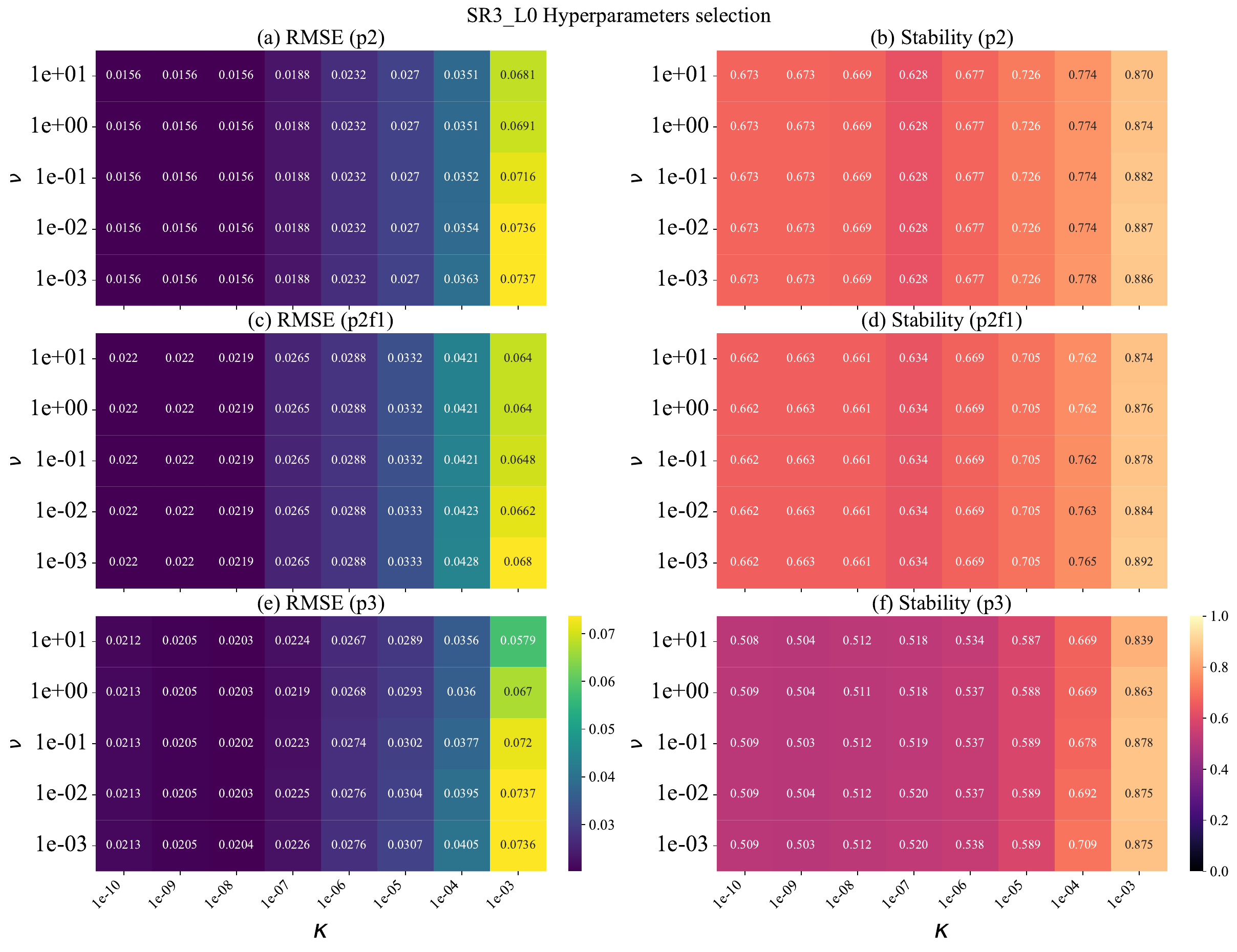}
\caption{\label{fig:heatmap_sr3_l0}Hyperparameter Sensitivity Analysis for SR3 ($L_0$). Heatmaps show mean stable RMSE and stability across sparsity parameter $\kappa$ and relaxation parameter $\nu$. RMSE remains nearly constant for small $\kappa$, while stability increase with $\kappa$.
}
\end{figure}

\begin{figure}
\centering
\includegraphics[width=1.1\columnwidth]{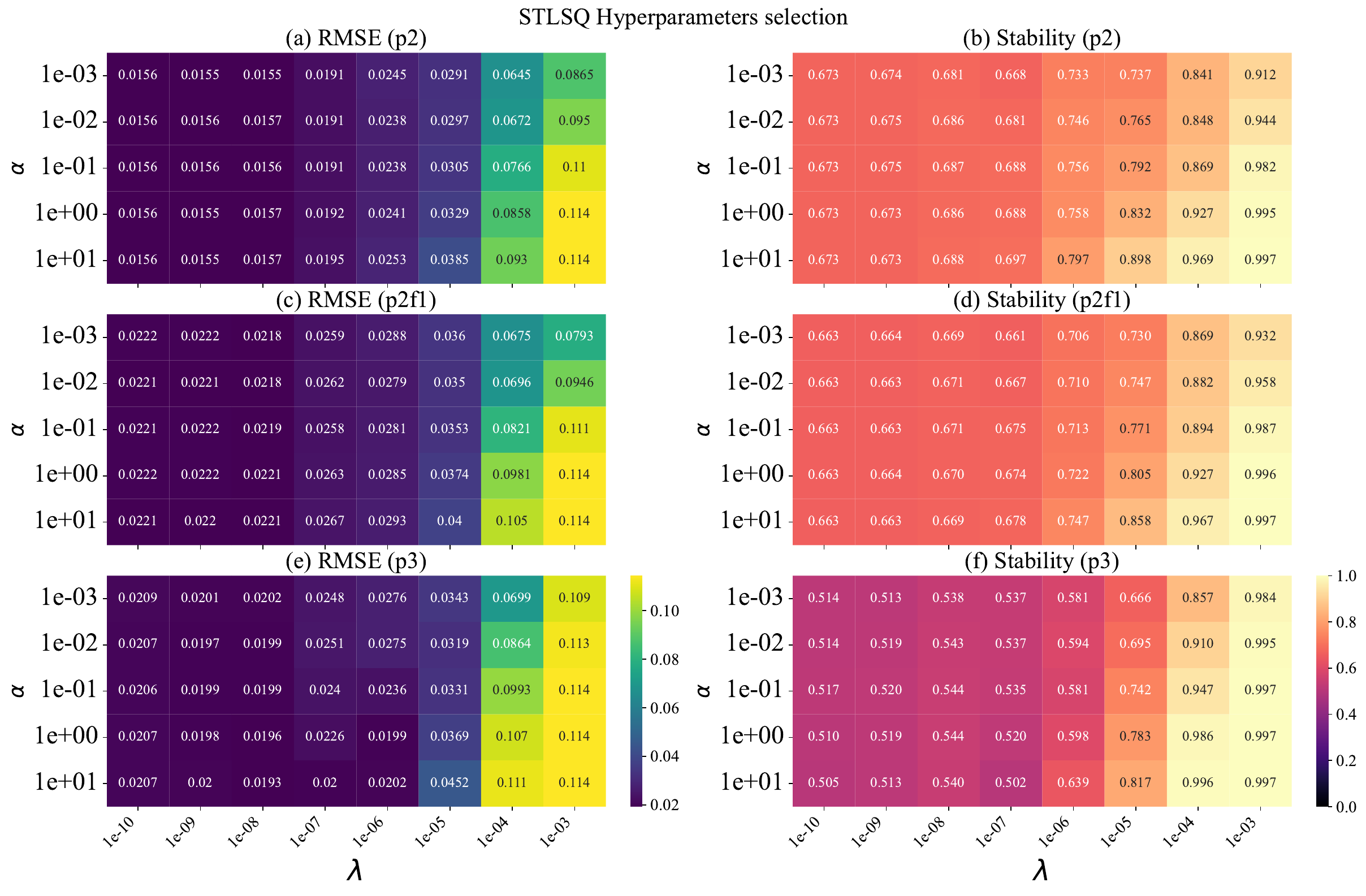}
\caption{\label{fig:heatmap_stlsq}Hyperparameter Sensitivity Analysis for STLSQ. Heatmaps show mean stable RMSE and stability across sparsity parameter $\lambda$ and ridgeparameter $\alpha$, highlighting a trade-off between accuracy and robustness.
}
\end{figure}

Figure~\ref{fig:heatmap_lasso}, ~\ref{fig:heatmap_sr3_l0} and ~\ref{fig:heatmap_stlsq} presents heatmaps of RMSE and stability fraction for the LASSO, SR3($L_0$), and STLSQ optimizer across the three library configurations, illustrating how hyperparameter sensitivity depends on model complexity.

For LASSO~(Figure~\ref{fig:heatmap_lasso}), the performance is primarily governed by $\alpha$.
For the quadratic model ($p2$) , RMSE remains nearly constant for samll $\alpha$, and increases sharply for larger $\alpha$. 
A similar pattern is observed for the $p2f1$ model. 
For the subic model ($p3$), the dependence on $\alpha$ becomes more pronounced, the RMSE achieves its minimum at $\alpha = 10^{-6}$. 
In contrast, the stability fraction increases monotonically with $\alpha$, indicating a clear trade-off between accuracy and dynamical robustness.

For SR3 with $L_0$ regularization~\ref{fig:heatmap_sr3_l0}, the performance is primarily governed by the sparsity parameter $\kappa$, while the relaxation parameter $\nu$ plays a secondary role.
Across all models, RMSE remains nearly invariant over a wide range of small $\kappa$ values. However, as $\kappa$ increases, RMSE increases rapidly. Stability, on the other hand, improves monotonically with $\kappa$, again reflecting a trade-off between accuracy and robutness.

For STLSQ~\ref{fig:heatmap_stlsq}, across all models, a board region of low RMSE is observed for small threshold values. However, as $\lambda$ increases beyond $10^{-5}$, RMSE rises sharply. 
The ridge parameter $\alpha$ has comparatively minor effect on RMSE but plays a stabilizing role: larger $\alpha$ values consistently improve stability fractions.

Across all three optimizers, a consistent pattern emerges: first, RMSE exhibits a broad plateau over several order of magnitude in sparsity-related parameter. Second, stability improves monotonically with stronger regularization. Finally, a trade-off between accuracy and robutness is present in all models.
Importantly, the selected hyperparametets used in the main results in Table~\ref{tab:ce-sk-comparison} lie well within these stable. This indicates that the reported results are not sensitive to small perturbations in parameter values, confirming the robustness of the model selection.


\subsection{Optimal Configurations from Grid Search}

The hyperparameter configurations that perform the best for each modelmand optimizer are summarized in Table~\ref{tab:best_configs}, identified by a systematic search of the grid in the CE data set. For each library and optimizer combination, we report the configuration achieving the lowest mean stable RMSE. 
For the quadratic model ($p2$), all optimizers achieve comparable performance, with minimal differences in RMSE and stability.
For the $p2f1$ model, LASSO achieves significantly higher stability (0.814) compared to other methods, indicating improved robustness.
In contrast, the more expressive cubic model ($p3$) achieves the lowest RMSE, but at the cost of reduced stability, highlighting the trade-off between model accuracy and dynamical robustness.

\begin{table}[h!]
\centering
\caption{Best hyperparameter configurations identified from grid search on the CE dataset. These values represent the optimal points for each combination. The parameters reported in Table 1 (Section~\ref{sec:results} are selected from within a robust performance range, i.e., a plateau in the hyperparameter space where RMSE remains nearly constant over several orders of magnitude, while stability stays within a comparable range for small regularization values before increasing monotonically, indicating that the selected parameters avoid regions of high sensitivity.}
\label{tab:best_configs}
\resizebox{0.7\textwidth}{!}{%
\begin{tabular}{lllcc}
\toprule
Model & Optimizer & Best Config & RMSE & Stability \\
\midrule
p2    & LASSO & $\alpha = 10^{-9}$ & \textbf{0.0155} & 0.673 \\
      & SR3 ($L_0$) & $\kappa = 10^{-10}, \nu = 1.0$ & 0.0156 & 0.673 \\
      & SR3 ($L_1$) & $\kappa = 10^{-10}, \nu = 1.0$ & 0.0156 & 0.673 \\
      & SR3 ($L_2$) & $\kappa = 10^{-5}, \nu = 0.001$ & 0.0155 & 0.673 \\
      & STLSQ & $\lambda = 10^{-9}, \alpha = 10$ & \textbf{0.0155} & 0.673 \\
\midrule
p2f1  & LASSO & $\alpha = 10^{-6}$ & \textbf{0.0160} & \textbf{0.814} \\
      & SR3 ($L_0$) & $\kappa = 10^{-8}, \nu = 0.001$ & 0.0219 & 0.661 \\
      & SR3 ($L_1$) & $\kappa = 10^{-8}, \nu = 0.1$ & 0.0219 & 0.661 \\
      & SR3 ($L_2$) & $\kappa = 10^{-6}, \nu = 0.1$ & 0.0219 & 0.667 \\
      & STLSQ & $\lambda = 10^{-8}, \alpha = 0.01$ & 0.0218 & 0.671 \\
\midrule
p3    & LASSO & $\alpha = 10^{-6}$ & \textbf{0.0101} & 0.586 \\
      & SR3 ($L_0$) & $\kappa = 10^{-8}, \nu = 0.1$ & 0.0202 & 0.512 \\
      & SR3 ($L_1$) & $\kappa = 10^{-7}, \nu = 0.1$ & 0.0180 & 0.512 \\
      & SR3 ($L_2$) & $\kappa = 10^{-7}, \nu = 0.001$ & 0.0140 & 0.541 \\
      & STLSQ & $\lambda = 10^{-8}, \alpha = 10$ & 0.0193 & 0.540 \\
\bottomrule
\end{tabular}
}
\end{table}

\end{document}